# ORBITAL PERIOD INCREASE IN ES CETI

Enrique de Miguel,[1,2] Joseph Patterson,[3] Jonathan Kemp,[4,*]
Gordon Myers,[5] Robert Rea,[6] Thomas Krajci,[7] Berto Monard,[8] and
Lewis M. Cook[9]

[1] *Departamento de Ciencias Integradas, Facultad de Ciencias Experimentales, Universidad de Huelva, 21070 Huelva, Spain*

[2] *CBA-Huelva, Observatorio del CIECEM, Parque Dunar, Matalascañas, 21760 Almonte, Huelva, Spain*

[3] *Department of Astronomy, Columbia University, 550 West 120th Street, New York, NY 10027, USA*

[4] *Mittelman Observatory, Middlebury College, Middlebury, VT 05753, USA*

[5] *CBA-San Mateo, 5 Inverness Way, Hillsborough, CA 94010, USA*

[6] *CBA-Nelson, Regent Lane Observatory, 8 Regent Lane, Richmond, Nelson 7020, New Zealand*

[7] *CBA-New Mexico, PO BOX 1351, Clodcroft, NM 88317, USA*

[8] *CBA-Kleinkaroo, Kleinkaroo Observatory, PO Box 281, Calitzdorp 6660, South Africa*

[9] *CBA-Concord, 1730 Helix Court, Concord, CA 94518, USA*

## ABSTRACT

We report a long-term study of the eclipse times in the 10-minute helium binary ES Ceti. The binary period increases rapidly, with $P/\dot{P} = 6.2 \times 10^6$ yr. This is consistent with the assumption that gravitational radiation (GR) drives the mass transfer, and appears to be the first dynamical evidence that GR is indeed the driver of evolution in this class of very old cataclysmic variables – the AM Canum Venaticorum stars.

Corresponding author: Enrique de Miguel
edmiguel63@gmail.com

* Visiting Astronomer, Cerro Tololo Inter-American Observatory, National Optical Astronomy Observatory, operated by the Association of Universities for Research in Astronomy (AURA) under a cooperative agreement with the National Science Foundation

## 1. INTRODUCTION

Among the cataclysmic variables (CVs), a few stars show a spectrum of nearly pure helium. With extremely short orbital periods (10-65 minutes), these binaries appear to be remnants of some process whereby the donor star has been previously stripped of all its hydrogen. Despite the exotic composition, and the likely exotic prior evolution, the present-day zoology of these stars follows the pattern characteristic of their hydrogen-rich cousins, with eruptive behavior essentially set by accretion rate. The stars of highest and lowest accretion rate (sometimes called "novalike variables") are fairly stable in brightness, while the intermediate regime shows sudden and frequent eruptions of 3-8 magnitudes in light ("helium dwarf novae").

These are the AM Canum Venaticorum stars, recently reviewed by Levitan et al. (2015) and Solheim (2010). For many years it was not a recognized class, since AM CVn was the only known member. Several classic studies of extraordinary foresight (Smak 1967; Paczynski 1967; Faulkner, Flannery & Warner 1972) suggested that AM CVn is a close binary with an accreting white dwarf, and with mass-transfer powered by the removal of angular momentum by gravitational radiation. The basic idea is that if the donor stars of otherwise ordinary CVs consist of helium, their orbital periods should be very short, since helium stars are much smaller than their hydrogen-rich cousins. But even the binary nature of AM CVn, the defining class member, remained in some doubt (Patterson et al. 1992). With the discovery and study of CR Boo (Wood et al. 1987) and V803 Cen (O'Donoghue et al. 1987; Patterson et al. 2000), it became increasingly clear that this was indeed a new class of CVs, and with the well-known suite of accretion-based behavior. A precise and stable binary period was found for AM CVn (Harvey et al. 1998; Skillman et al. 1999), and the puzzling array of photometric periods in this and other stars became readily understood as superhumps (precessional sidebands of the orbital frequency, analogous to those in the Earth-Moon-Sun system). In recent years the roster has grown to $\sim$40 confirmed and probable members (Levitan et al. 2015; Warner 2015).

The question remains: what drives the mass-transfer in these binaries? Since Smak's landmark 1967 paper, the common assumption has been: gravitational radiation (GR), which powers mass transfer by removing angular momentum. But there has never been any evidence to support, or for that matter deny, that hypothesis. In principle, the observed rate of orbital-period change can test the hypothesis. But this test is not feasible for most members of the class, since observations of sufficient precision, over a sufficiently long time, have not yet been obtained. In this paper we report a study which reveals that rate of period change for one important member of the class, ES Ceti.

## 2. THE STAR, AND THE OBSERVATIONS

ES Ceti was discovered as an ultraviolet-bright object in the Kiso survey (Kondo et al. 1984), with a helium emission-line spectrum (Wegner et al. 1987). Warner &

Woudt (2002) found a persistent photometric period of 620.2 s, which appeared to be an *orbital* period and therefore strongly suggested membership in the AM CVn class. Studies on a longer baseline [Espaillat et al. (2005), hereafter EPWW; Copperwheat et al. (2011), hereafter C2011] established a precise period of 0.00717837598(3) d = 620.211685(3) s, with $|\dot{P}| < 10^{-11}$.

We obtained new time-series photometry with the globally distributed telescopes of the Center for Backyard Astronomy (CBA). Skillman & Patterson (1993) and de Miguel et al. (2016) describe the methods and observing stations of the network. Typically we observed in unfiltered light, at 10-40 s time resolution for 2-5 hours each night. A total of seven CBA stations were involved in the ES Ceti campaign. The data were differential photometry with respect to a nearby star. The primary comparison stars used were UCAC4 404-002281 ($V = 13.03$) and UCAC4 403-002351 ($V = 15.27$), with occasional use of other comparison stars. They were systematically checked and were found to show no signs of variability. A few time series affected by adverse conditions or low signal-to-noise ratio were rejected. Including the work of EPWW, the data comprised 452 hours over 145 nights in 2003-2017.

The nightly light curves were always very similar to that shown by EPWW. We summed synchronously (with respect to the 620 s period) over each night's time series, and fit a quadratic to the region around minimum light. This yielded a mean timing of minimum light each night, or each several nights when the runs were short and/or of less quality. Thus, some timings arise from "clusters" of nights; this helped to insure no large variations in data quality. The resultant set of timings show a scatter of less than 20 s. The waveform varied slightly from night to night, but the star never strayed more than 0.1 magnitudes from its mean brightness of $V = 16.95$.

This was also the procedure followed in our earlier paper (Table 2 of EPWW). The timings of minima reported in Table 1 of C2011 were derived in a slightly different way: by fitting a mean waveform. Since many of our time series were also analyzed and published by C2011, we were able to measure the time difference between their measurement and ours to be consistently $37 \pm 6$ s. This was somewhat puzzling, because the difference in time stamps of their system (BJD$_{\rm TDB}$) and ours (HJD) was nearly constant at $65 \pm 3$ s during this interval. Probably the difference lies in method of light-curve measurement (C2011 fit the entire light curve, whereas we fit only the minimum; variability in the waveform, discussed by EPWW and C2011, renders the issue debatable). Anyway, we subtracted 37 s from the 14 timings of C2011 which were not included in EPWW or the present paper, and include them in the total collection. In addition to those timings, we here present 68 additional timings during 2006-2017. These HJD timings are reported in Table 1.[1] Combining these with the adjusted C2011 times, we obtain a total of 130 timings during 2001-2017.

**Table 1.** New timings of minimum light (HJD 2,450,000 +)

| | | | |
|---|---|---|---|
| 4060.70821 | 4063.70865 | 4074.79950 | 4102.57225 |
| 4104.56813 | 4449.79810 | 4463.57301 | 4479.58082 |
| 4497.58419 | 4766.92428 | 4768.91248 | 4769.89585 |
| 4770.92237 | 4777.92839 | 4778.86853 | 4779.82343 |
| 4782.87431 | 4784.86310 | 4787.87082 | 4796.87935 |
| 5137.91682 | 5141.90119 | 5142.89882 | 5147.91672 |
| 5195.71736 | 5197.59819 | 5504.63175 | 5505.62958 |
| 5514.60239 | 5519.59180 | 5850.67295 | 5931.59465 |
| 6323.29062 | 6570.7867 | 6574.35458 | 6575.33799 |
| 6579.35077 | 6594.41095 | 6602.30717 | 6609.34927 |
| 6618.40134 | 6674.56467 | 6903.44754 | 6960.10675 |
| 6960.39429 | 6962.41089 | 6964.42813 | 6971.44856 |
| 6972.93433 | 7258.57670 | 7273.58677 | 7275.22344 |
| 7277.17599 | 7278.10209 | 7279.20743 | 7314.95585 |
| 7318.92541 | 7627.15837 | 7628.16327 | 7629.18276 |
| 7630.08014 | 7636.06668 | 7688.31092 | 7716.92462 |
| 7717.92222 | 7719.27845 | 7720.28350 | 7722.27923 |

In Figure 1 we show an O-C diagram relative to the average period of 0.00717838 d. The drifting phase is satisfactorily fit by the parabolic[2] ephemeris

$$\begin{aligned}\text{Minimum light} = &\text{HJD } 2{,}452{,}201.3943(1)\\ &+ 0.007178371(2)\ E\\ &+ 1.14(4) \times 10^{-14}\ E^2,\end{aligned} \quad (1)$$

corresponding to $\dot{P} = 3.2(1) \times 10^{-12}$ and hence a period-increase timescale $P/\dot{P} = 6.2(2) \times 10^6$ yr. ES Cet is increasing its orbital period, as all properly-credentialed AM CVn stars should (see discussion below).

### 3. INTERPRETATION

In the absence of mass transfer, all close binaries should evolve to *shorter* period, driven by angular momentum loss – either by gravitational radiation, or by some more powerful process, **such as magnetic braking** (Verbunt & Zwaan 1981). Mass transfer, however, produces a period increase whenever mass flows from the low-mass to the high-mass component (as is nearly always the case in CVs). The battle between these effects produces a net period change. For the typical situation in CVs (mass of the accretor $M_1 \sim 0.8 M_\odot$, and donor mass $M_2 = 0.1 - 0.5 M_\odot$), angular-momentum loss ordinarily wins, with $P_{\rm orb}$ decreasing and radius $R_2$ decreasing as the donor evolves to lower mass (Rappaport et al. 1982; Patterson 1984; Knigge, Baraffe & Patterson 2011). But this depends on the properties of the donor, which must decrease in radius with decreasing mass (if $P_{\rm orb}$ is decreasing). Solar-type main-sequence stars easily satisfy this ($R \sim M^{0.9}$), and so do the donors in H-rich CVs [$R \sim M^{0.5}$; Knigge (2006); Patterson et al. (2005)].

---

[1] Strictly speaking, HJD is an inferior time system, since it does not correct for the motion of the solar-system barycenter (±3 s), and contains discontinuities (leap seconds). However, these effects are negligible for this study, and we prefer HJD as it is still the more commonly used standard in astronomical photometry.

[2] It might appear that a more complex (cubic?) fit is required, since nearly all points in the range $E = 100{,}000 - 300{,}000$ are above the fitted curve. But nearly all these points are non-CBA timings adjusted by subtracting 37 s from the C2011 table. This correction is uncertain, as discussed above. Had we subtracted 44 s, the points would straddle the curve. With differing techniques of measurement, this seems like an insignificant difference.

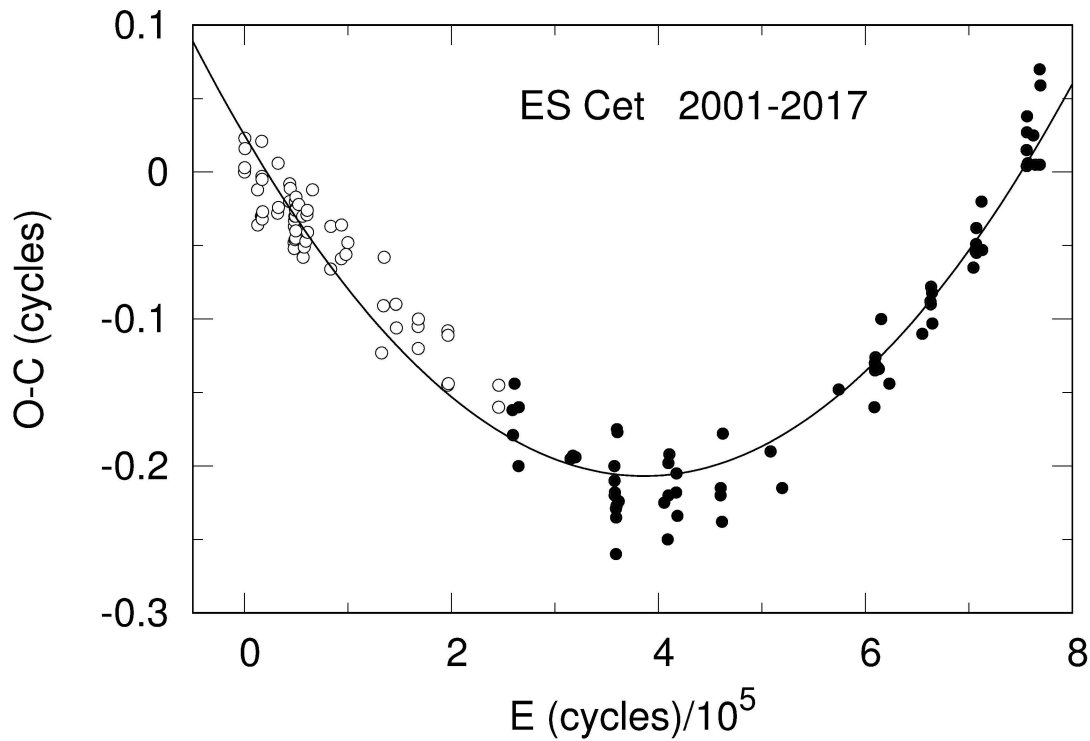

**Figure 1.** O-C diagram of the timings of minimum light, relative to the test ephemeris: HJD $2,452,201.3941 + 0.00717838E$. **Filled symbols indicate new measurements from Table 1. Open symbols indicate measurements from EPWW and C2011.** The fitted parabola corresponds to Eq. 1, which indicates a period increase with $P/\dot{P} = 6.2(2) \times 10^6$ yr.

But little is known about the helium-rich donors in AM CVn stars. A common name for the class ("double degenerates") implies that the donor's radius scales inversely with mass, with $R \sim M^{-0.33}$ at low mass (the original Chandrasekhar relation). This relation is for cold degenerate matter. Deloye, Bildstein & Nelemans (2005) (hereafter DBN) find $R \sim M^{-0.28}$ for the low-mass donor stars, assumed degenerate, in AM CVn binaries.

We also know that the donor in ES Cet fills its Roche surface. This yields a $P - \rho^{1/2}$ relation for the donor, which, for the DBN degenerate mass-radius relation, implies a donor mass $M_2 = 0.062 M_\odot$. For the fully conservative case (pure mass transfer, with no mass or angular momentum loss from the binary), the period change is given by

$$\dot{P}/P = 3(1-q)(\dot{M}/M_2), \qquad (2)$$

where $\dot{M} \equiv -\dot{M}_2$ is the mass-transfer rate from the donor, and $q = M_2/M_1$ is the mass ratio. Since we know $\dot{P}$, $P$ and $M_2$, we can solve this for $\dot{M}$ and obtain

$$\dot{M} = 3.4 \times 10^{-9} (M_1/M_\odot)^{-0.1} \, M_\odot \text{yr}^{-1}. \qquad (3)$$

Of course, angular-momentum loss, at least by GR, must also be present, and by itself leads to a period decrease. Therefore $\dot{M}$ must be even higher, to account for the observed period increase. The assumption of a degenerate secondary (double degenerate) is also open to doubt.

For ES Cet we have measured $\dot{P}/P = 5.3 \times 10^{-15}$ s$^{-1}$ and this is related to $\dot{J}/J$ by Eq. (1) of C2011:

$$\frac{\dot{P}}{P} = 3\left[\frac{\dot{J}}{J} - (1-q)\frac{\dot{M}_2}{M_2}\right]. \qquad (4)$$

The angular momentum loss driving the mass transfer is assumed to be due to GR:

$$\frac{\dot{J}}{J} = -\frac{32G}{5c^5}\frac{M_1 M_2 (M_1 + M_2)}{a^4}. \qquad (5)$$

where $a$ is the separation between the binary components. Applying Kepler's Third Law and the measured $\dot{P}/P$, we can solve for $\dot{M}$ as a function of $M_1$. For a degenerate secondary, the result is

$$\dot{M}_1 = -\dot{M}_2 = 9.0 \times 10^{-9} (M_1/M_\odot)^{0.3} \, M_\odot \text{yr}^{-1}. \qquad (6)$$

A semi-degenerate secondary would be somewhat more massive, and that dependence scales like:

$$\dot{M} = 9.0 \times 10^{-9} (M_2/0.062 M_\odot)^{1.7} \, M_\odot \text{yr}^{-1}. \qquad (7)$$

Among short-period CVs, this is a pretty high accretion rate. Except for remnants of recent novae, the accretion rates – inferred from luminosity – in short-period H-rich CVs are generally near $10^{-10} M_\odot \text{yr}^{-1}$. Is it reasonable to believe that ES Cet could be accreting at a rate 100x greater?

Yes, it is quite reasonable. With its extremely blue colors and very strong He II emission, the star is clearly dominated by very high temperature gas, signifying high $\dot{M}$ if it originates from a disk – which is likely, since the emission lines show the doubled profile characteristic of a disk. This evidence was studied by EPWW, who estimated an accretion rate

$$\dot{M}_1 = (2-4) \times 10^{-9} (M_1/0.7 M_\odot)^{-1.7} (d/350\,\mathrm{pc})^2 \, M_\odot \mathrm{yr}^{-1}. \qquad (8)$$

With small adjustments in distance or $M_1$, this is within hailing distance of Eq. 6, the estimate from period-change evidence. Of course, it is also possible that some fraction of the mass lost by the donor escapes from the binary, never accreting onto the primary. This would carry off angular momentum, and thereby decrease $P_\mathrm{orb}$; so the mass transfer rate would have to be even higher (in order to be consistent with the rate of period increase).

It also explains why ES Cet does not show dwarf-nova outbursts: because the temperature is too high, and the disk remains in a permanently hot state. Figure 9 of EPWW illustrates the run of $\dot{M}$ with $P_\mathrm{orb}$ for all the AM CVn stars known at the time (2005). With an accretion rate near $10^{-8} M_\odot \mathrm{yr}^{-1}$, ES Cet is comfortably in the novalike regime, with no outbursts expected. To express it differently, the star is in *permanent* outburst.

These results also appear to be consistent with theory. In Figure 2 of DBN, an accretion rate around $10^{-8} M_\odot \mathrm{yr}^{-1}$ is expected for a 10-minute binary. And in view of the sensitivity to $M_2$ in Eq. 7, it is probable that the donor mass does not much exceed $0.062 M_\odot$ (the value for full degeneracy).

## 4. DISCUSSION

This appears to be the first successful test of the GR hypothesis, based on $\dot{P}$, for any cataclysmic variable. Among the majority species, the H-rich CVs, a few stars show small $\dot{P}$ effects, but with an accumulated wandering in phase by no more than ∼0.01 cycles, and sometimes of alternating sign. These are of unknown origin, but may be due to magnetic cycles in the secondary (Warner 1988; Applegate 1992; Richman et al. 1994; Parsons et al. 2010; Han et al. 2017).

A more important and interesting test of the GR hypothesis for H-rich CVs lies in the "period minimum" test – for CVs as a class, not for any particular star. For donor stars of solar composition and with GR as a driver, the binary periods should decrease down to 67 minutes, and then slowly creep back up towards longer period (as a "period bouncer"). The observed period minimum is in the range 77-80 minutes – probably with an intrinsic spread, but quite distant from the 67 min predicted by GR. Knigge, Baraffe & Patterson (2011) studied this discrepancy and concluded that if the driver is angular-momentum loss, then it must exceed the loss rate due to GR by a factor of 3.

Among the *bona fide* AM CVn stars, there has been no useful $\dot{P}$ data. For such a measurement, **we need a star with a fairly precise signature of orbital phase [e.g. eclipsing systems, such as YZ LMi (Szypryt et al. 2014), or Gaia14aae (Campbell et al. 2015)]. And we need to track it for 10 years** or more (possibly much more if the orbital period is long). With its short period, very low flickering, long baseline of measurement, and accessibility to small telescopes, ES Cet is presently the only star which qualifies.[3] Nor is there a minimum-period known for AM CVn stars; so that test is not available.

There are $\dot{P}$ measurements for 2 AM CVn *candidates*: HM Cnc ($P = 321$ s) and V407 Vul = RX1914+24 ($P = 569$ s). The former period is very likely orbital (Roelofs 2010), while the latter's candidacy is murkier, since the star's spectrum does not even show helium. These two periods are **decreasing**, at rates which could plausibly be interpreted as due to angular-momentum loss via GR (Strohmayer 2004, 2005). But since the periods are decreasing, the donors must be either detached, or contracting as they lose mass – inconsistent with a mass-radius relation for a degenerate star. Deloye & Taam (2006) [see also Kaplan et al. (2012)] suggest that these two stars can be in a short-lived phase of mass transfer, when the two WDs, rapidly inspiralling under the action of GR (thus explaining the period decrease) just reach Roche-lobe contact and begin mass transfer (thus explaining the accretion luminosity). It's a delicate balancing act, and can only last for a short time.

Finally, it is possible that the curvature in Figure 1 arises merely from the reflex motion of ES Cet around a very distant third body. The putative orbital period would be around 30 years or more, and a late M dwarf ∼10 AU distant could be an acceptable fit. Weighing against this possibility is the absence of such effects in any other CV studied to date ... and perhaps the cruelty of a commonplace and unwanted star reproducing the effect of GR with good accuracy.

In summary, our 15-year study of the 10-minute orbital wave in ES Ceti reveals a period increase with $P/\dot{P} = 6.2(2) \times 10^6$ yr. This is consistent with the assumption that the mass transfer is driven by angular momentum loss from gravitational radiation, and appears to be the first AM CVn star which tests (and satisfies) that expectation. Further tests for long-term period changes in this class are warmly recommended.

We are grateful to NSF through grant AST-1615456, and to the Mount Cuba Astronomical Foundation, for their support of our research.

---

[3] We note that Szypryt et al. (2014) have reported a positive $\dot{P}$ in the eclipsing AM CVn star YZ LMi. But this is based just on three epochs (2006, 2009, and 2012) and needs confirmation.